\documentclass[journal]{IEEEtran}
\usepackage{cite}
\usepackage{graphicx}
\ifCLASSINFOpdf
\graphicspath{{pics/}}
\DeclareGraphicsExtensions{.pdf,.jpeg,.png}
\else
\fi
\usepackage{hyperref}
\usepackage{makecell}
\usepackage{url}
\usepackage{booktabs}
\usepackage{multirow}
\usepackage{subfigure}
\usepackage{amsmath}
\usepackage{amssymb}
\usepackage{color}
\usepackage{enumitem}
 \usepackage{mathrsfs}
\usepackage{siunitx}
\newcommand{\iu}{\mathrm{i}\mkern1mu}

\begin{document}

\title{Spatial and Temporal Characterization of Living Mycelium through Dispersion Analysis}
\author{Bao Zhao\IEEEauthorrefmark{1},~\IEEEmembership{Member,~IEEE}, Sophia Ganzeboom\IEEEauthorrefmark{1}, Marcus Haywood-Alexander, Eleni Chatzi,~\IEEEmembership{Member,~IEEE}, and Vasilis Dertimanis
\thanks{The authors acknowledge the support of the SFA-AM (Strategic focus area Advanced Manufacturing) Initiative under the project ALIVE, Advanced Engineering with Living Materials.
(\textit{Corresponding authors: Bao Zhao and Sophia Ganzeboom}.)}
\thanks{The authors are with the Department of Civil, Environmental and Geomatic Engineering, ETH Zürich, 8093 Zürich, Switzerland (e-mail: bao\_zhao@outlook.com; sophia.ganzeboom@ibk.baug.ethz.ch; Marcus.Haywood-Alexander@ibk.baug.ethz.ch; chatzi@ibk.baug.ethz.ch; v.derti@ibk.baug.ethz.ch)}
\thanks{Bao Zhao is also with the Department of Civil and Environmental Engineering, Hong Kong Polytechnic University, Kowloon, Hong Kong}
\thanks{\IEEEauthorrefmark{1} These authors contribute equally to this paper.}
\thanks{\textit{Preprint}}
}
\maketitle

\begin{abstract}
Mycelium, a natural and sustainable material, possesses unique electrical, mechanical, and biological properties that make it a promising candidate for biosensor applications. These properties include its ability to conduct electrical signals, respond to external stimuli such as humidity and mechanical stress, and grow integrally within structures to form a natural network. Such characteristics suggest its potential for integration into self-sensing systems to monitor vibrations, deformations, and environmental conditions in buildings and infrastructure. To understand the output voltage generated by these biomaterials in response to an applied electrical input, it is essential to characterize their spatial and temporal properties. This study introduces an electrical impedance network model to describe signal transmission through mycelium. In combination with the inhomogeneous wave correlation (IWC) method—commonly used in elastic wave propagation—we demonstrate, for the first time, the dispersion behavior of living mycelium both theoretically and experimentally. We reveal the frequency-dependent and spatial attenuation of electrical signals in living, dehydrated, and rehydrated mycelium, emphasizing the critical role of humidity in enabling effective signal sensing. Furthermore, dispersion analysis is used to assess the homogeneity of mycelium, underscoring its feasibility as a living, green sensing material. This research lays the groundwork for innovative applications of mycelium in sustainable structural health monitoring. 
\end{abstract}


\section{Introduction}
\label{sec:intro}

The increased demand for sustainable and biodegradable materials has sparked interest in the adoption of biobased materials for technological applications. The mycelium, the vegetative part of the fungus, is one of the main organisms studied as a sustainable material. The fungus kingdom spans an enormous diversity of taxa, including organisms such as yeast, mold, and oyster mushrooms, which thrive across an extraordinary range of environmental conditions due to their remarkable adaptive capacities \cite{coleineFungiAreKey2022}. These survival mechanisms have already demonstrated significant potential in various biotechnological applications  \cite{buffi2025electrical}.
Recently, fungi have gained attention in further science fields as a promising multipurpose substance. The cultivation of mycelium that grows beneath the soil or inside a substrate has opened up a new way to produce biodegradable and sustainable materials with diverse usage in various sectors, ranging from packaging materials \cite{mananSynthesisApplicationsFungal2021b}, to insulation materials \cite{schrittSpentMushroomSubstrate2021}, to textiles \cite{jonesLeatherlikeMaterialBiofabrication2021}. This versatility of using mycelium can be attributed to its unique ability to break down organic matter and transform it into complex structures. The combination of the large variety of fungi with their unique characteristics and the type of substrate in which the mycelium grows has led to numerous materials for different fields with different properties and characteristics.

While the production processes for mycelium-derived materials vary significantly, a common characteristic involves inactivating the mycelium through heating and dehydration \cite{jones2017mycelium,jones2020engineered}.
However, recent research efforts have shifted towards maintaining mycelium in a living and active state, enabling novel applications \cite{adamatzky2023fungal,mishra2024sensorimotor,schyck2024harnessing}. Moreover, the inherent ability of living mycelium to adapt to and sense its environment presents opportunities for developing materials with intrinsic self-repair mechanisms or environmental responsiveness, such as detecting structural or environmental changes, which reveal damage or deterioration. Central to this capability are the unique electrical properties of mycelium.

Given the intrinsic network of fungal hyphae, ions generated by fungal biochemical activities can support electrical conduction in aqueous environments in vivo \cite{gow1984transhyphal,gow1995electric}. In 1962, Slayman et al. \cite{slayman1962measurement,slayman1976action} first discovered that fungi exhibit non-trivial electrical activity. When measuring their intracellular electric potential, it was observed that the tested fungi exhibited action potential-like spikes lasting between 30 seconds and 2 minutes. Since then, the team of Adamatzky et al. has conducted extensive investigations into this phenomenon \cite{adamatzky2018spiking,adamatzkyFungalElectronics2021,adamatzky2022language,adamatzky2023fungal}, including the study by Dehshibi et al. \cite{dehshibiElectricalActivityFungi2021}, which proposed various processing methods for characterizing fungal electrical behavior. Recently, the promising trend to utilize fungi or mycelium as smart sensors has gained significant attention \cite{danninger2022myceliotronics,mishra2024sensorimotor}. Various instances of electrical behaviors of fungi have been reported, such as spontaneous spiking behaviors \cite{gow1984transhyphal,adamatzky2018spiking,ganzeboom2024fungal}, sensitivity to moisture change \cite{phillips2023electrical}, chemical stimulation \cite{olsson1995action}, physical loading \cite{nikolaidou2023responsive}, electrical stimulation \cite{mcgillivray1986applied,phillips2024electrical}, and even light radiation \cite{horwitz1984electric,mishra2024sensorimotor}. Most existing research works mentioned above focus on studying the electrical behavior of fungal mycelium in a localized manner \cite{fukasawa2024electrical,jones2024fungal,phillips2024electrical}. While used as sensors or transmission media, the spatial geometry of the fungi will also influence the propagation of the electrical signal. The frequency response function with different distances between input and output electrodes may vary dramatically. This spatially induced propagation characteristic has been well studied in the classical transmission line problem, considering both influences of frequency and wave number for electromagnetic wave propagation in finite conducting materials \cite{pozar2021microwave}.
However, how the electrical signal propagates in the fungal material with a certain length or geometry still remains an open question.

A robust model that is able to adequately describe the mycelium's electrical behavior ought to capture the equivalent impedance or dielectric properties of the tested sample \cite{stupin2021bioimpedance}, especially when the parameters are frequency-dependent due to different polarization mechanisms under the external electrical field \cite{schwan1992linear}. Early attempts in this direction mainly focused on studying the electrical impedance spectroscopy of animal or human tissues \cite{dean2008electrical}. Built on the frequency-dependent dispersion and permittivity studies of human tissues, body area communications with human skin \cite{seyedi2013survey,callejon2011study} and biomedical imaging with dielectric spectroscopy \cite{wu2021electrical} have been proposed to characterize biomaterials for sensing and medical purposes. While the investigation of fungal mycelium as a smart sensor could be cast in a similar track, existing studies of its dielectric or impedance properties mostly report on observational findings without a rigorous derivation of a model that describes the associated phenomenon. A recent attempt toward such a rigorous model has been delivered by Phillips et al. \cite{phillips2024electrical}, who proposes a discrete series resistor-capacitor (RC) impedance model for modeling the electrical response of fungi. This lumped-parameter model captures the low-pass characteristic of mycelium material when an external electrical field is applied. This model captures the general dielectric nature of fungal material. However, the model is fitted based on locally measured response signals without consideration of the spatial geometry of the mycelium sample, thus hindering an extensive spatial characterization.

The spatial characterization requires more than the frequency response function of a single position on the mycelium sample. Multiple frequency responses along a sampled line are needed to construct the resolution in the spatial domain, which leads to the concept of the dispersion relationship that governs the spatial and temporal wave propagation. The dispersion relationship is essential for media and sensors interacting via wave propagation. It has been extensively studied and well understood in other applications of materials that can serve for sensing purposes. Taking the pertinent example of metamaterials \cite{craster2012acoustic,de2024localized,zhao2024nonlinear}, the wave propagation in these systems is governed by architected dispersion relationships, which allow the materialization of local phenomena that can be exploited for energy conversion and sensing functions \cite{zhao2023circuit}. When now turning to studying fungal mycelium as a sensor, it becomes important to study its electrical wave and signal propagation, and possibly associate such behavior with its function as a living sensor. In this work, we evaluate the dispersion relationship governing fungal mycelium samples.
We associate the impedance model with the plane wave propagation in the mycelium material, offering, for the first time, an analytical expression to characterize mycelium samples. We further propose the use of the inhomogeneous wave correlation (IWC) method \cite{van2018measuring,zhao2022graded} to characterize the experimental dispersion relationship and isotropy of the sensed mycelium.

The remainder of the paper is organized as follows: Section~\ref{sec:material} overviews the experimental protocol put in place for cultivating the mycelium samples, along with the electrical voltage measurement setup. Section~\ref{sec:method} proposes the relationship that defines the impedance network of fungal mycelium, based on the transmission line model and its analytical and experimental dispersion relationships for electrical signal propagation within a finite mycelium sample. The experimental results are presented in Section~\ref{sec:exp}. Section~\ref{sec:dis} discusses the ionic conduction of the mycelium material and provides a comparison with existing research on electrical activities of fungal mycelium.  Section~\ref{sec:con} concludes this paper.

\section{Materials Preparation}
\label{sec:material}

\subsection{Materials}

\begin{figure}[!t]
\centering
\includegraphics[width=1\columnwidth,page=1]{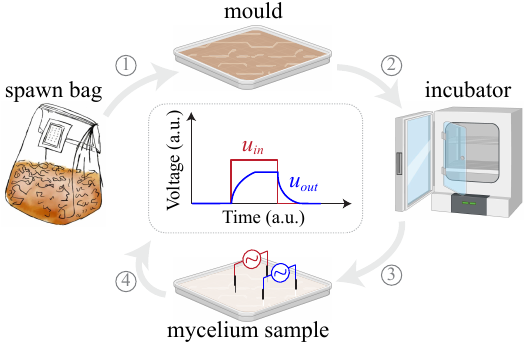}
\caption{Preparation of the mycelium sample.}
\label{fig:pre}
\end{figure}

\subsection{Growth Media and Mycelium Culture}

The preparation of the mycelium sample is illustrated in Fig. \ref{fig:pre}.
In all experiments reported in this study, the substrate material consisted of commercially available spawn bags from Grown.bio \cite{grownInnovativeMycelium}, which combine organic agricultural waste with mycelium (Pleurotus ostreatus). As shown in Fig. \ref{fig:pre}, the substrate was first mixed with flour as a nutrient source (30 g of flour per kilogram of substrate), and the composite was placed into a 30$\times$40 cm mold.
The mold was sealed and stored in incubators at a temperature of 25~$^o$C at  60\% humidity for 4 days during the growth of the mycelium composite.
Once the mycelium colonized the substrate, forming a uniform white skin, the sample was considered ready for testing.

\subsection{Electrical Voltage Input and Output}

The experimental campaign was initiated to estimate the sample's step response, which captures the local voltage evolution in the time domain of the mycelium with the input voltage changing from zero to a high value. As Fig.~\ref{fig:pre} illustrates, a square wave voltage $u_{in}$ was applied via a pair of electrodes across the midpoint of the sample, and the response $u_{out}$ was measured via another pair of electrodes close to the input electrodes. Based on the measured voltage waveforms, a series resistor-capacitor induced low-pass-filter response is clearly captured, revealing the dielectric nature of the mycelium sample. This time domain behavior has also been reported in Phillips et al. \cite{phillips2024electrical}, where the authors utilized a series RC model to simulate the local electrical behavior of the mycelium. However, the distance between the input and output electrodes also influences the measured voltage response. This indicates a dependency of the output voltage on the distance along the measured direction and its frequency dependence, which is an important consideration when wishing to deploy mycelium as a living sensor using samples confined by finite geometry.

\begin{figure*}
\centering
\includegraphics[width=2\columnwidth,page=2]{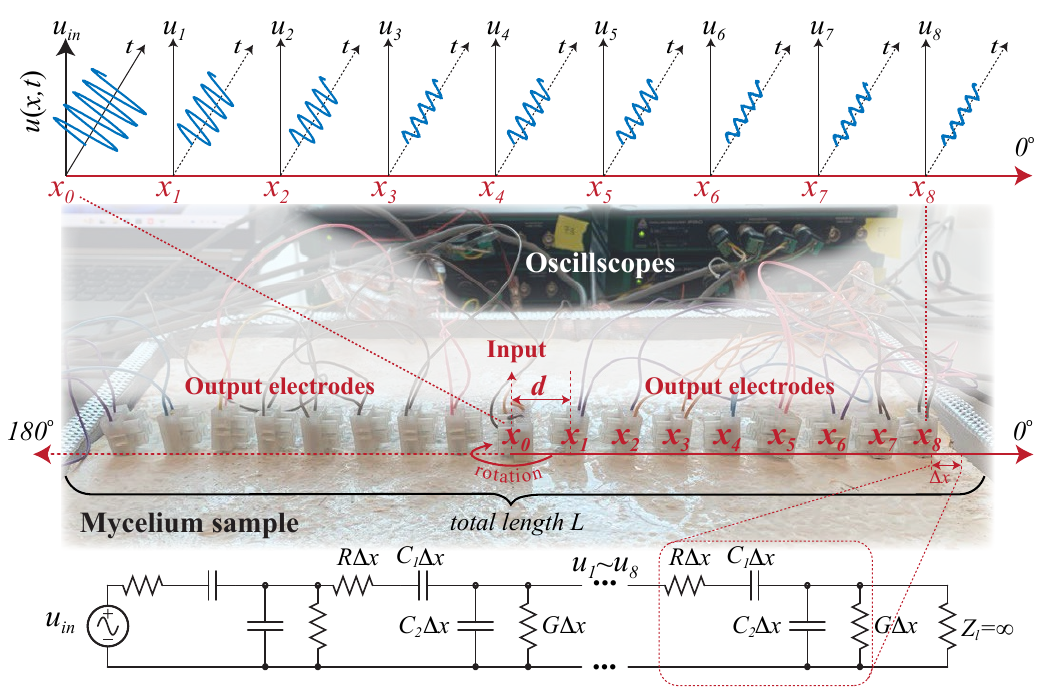}
\caption{Experimental setup and the impedance model for the fungal mycelium. Multiple pairs of differential electrodes are placed with a fixed distance $d$ along a certain electrical signal propagation direction. The output voltages from the sweep input $u_{in}$ are collected and noted as $u_1$ to $u_8$. Along the propagation direction, each infinitesimal segment of the mycelium sample is modeled as a two-port impedance network. The ending part of this transmission line is assumed to be in open circuit condition with an infinite load $Z_l=\infty$.}
\label{fig:setup}
\end{figure*}

In order to quantify the spatiotemporal electrical attributes of the mycelium, the sample was actuated using a grid-like sensor setup. A single pair of electrodes was used as the actuating input voltage source. In contrast, the voltage response was captured using 16 further electrode pairs, placed at different positions and angles within the sample. The spacing was held approximately equal to 2.5 cm. As shown in Fig. \ref{fig:setup}, the input electrode pair is placed at the midpoint of the sample, while the response is measured sequentially at various angles, starting from $0^\circ$ and varying at an increment of $30^\circ$. The mycelium sample is then actuated with an input voltage on one end of the sample, with the response measured with all eight electrode pairs in one direction to capture the spatial response induced by the propagation of the actuating voltage across the sample.

To reveal the spatiotemporal characterization of the mycelium across a broadband width, a broadband sweep signal ranging from 1 Hz to 500 kHz is utilized as the input voltage signal. Four high-resolution digital oscilloscopes (Analog Discovery Pro 3000, ADP345, Digilent) with an analog input impedance of 1 MOhm and 14-bit resolution were coupled together. Their triggers are synchronized to be used as a signal generator and a 16-channel data logger for sending sweep voltage excitations and recording voltage signals.
The differential voltage was recorded through iridium-coated stainless steel subdermal needle probes (Specs Medica, Italy).
Due to the oscilloscope's limited buffer size, the sweep signal is separated into 10 sections, each covering a frequency range of 50 kHz. The sampling frequency is set to 6.25 MHz to satisfy the Nyquist sampling criteria. All devices are commonly grounded to avoid ground noise.

\section{Methodologies}
\label{sec:method}
\subsection{Impedance Modeling}
\label{sec:impedance}

The electrical measurement is initially conducted along a certain line of the sample, as shown in Fig.~\ref{fig:setup}. Taking the case of $0^\circ$ direction as an example, the input electrical voltage excitation $u_{in}$ is placed at the geometric center ($x_0$) of the mycelium material, and the output differential probes are distributed along the $0^\circ$ direction of the material with a gap distance $d$. The electrical output voltages $u_1$ to $u_8$, at positions $x_1$ to $x_8$, represent the response of the sample along the $0^\circ$ direction, and are used to analyze the electrical properties of the mycelium material along this direction.

The permittivity of biological materials can be frequency-dependent due to multiple reasons, such as the material micro-structure, polar molecules and ions, or others. This implies that the capacitance of fungal mycelium could be frequency-dependent, as is the case for biological tissues \cite{foster2019dielectric}. However, here we only focus on the low to medium frequency ranges (lower than 1 MHz) due to the limited buffer size and bandwidth of the digital oscilloscopes. We further place the assumption of homogeneity of the mycelium material at the macro level (intracellular level) when sampled under low-frequency input signals. This assumption is further verified in Sec.~\ref{sec:exp_dis}. To model the electrical behavior of the mycelium under these conditions, we turn to the transmission line theorem \cite{pozar2021microwave}, which provides a well-established framework for analyzing signal propagation in distributed-parameter systems. This approach allows us to construct an equivalent impedance network, as illustrated in Fig.~\ref{fig:setup}.
The impedance model approximates the mycelium sample as an infinite series of two-port networks which relate the input and output voltages and currents, each representing an infinitesimally short portion of the mycelium material with components:
\begin{enumerate}
    \item $R$: Resistance per unit length to the propagation of the signal in the longitudinal direction.
    \item $C_1$ and $C_2$: Capacitance per unit length in the longitudinal and transverse directions, respectively. They characterize the dielectric properties of living mycelium, describing the ability to store electrical charge and energy \cite{mcadams1995tissue}.
    \item $G$: Conductance per unit length in the transverse direction, which describes the dielectric loss \cite{danninger2022myceliotronics} of the mycelium material through a shunting resistance $1/G$ that allows a small leakage current to flow across the dielectric, representing energy dissipation within the material.
\end{enumerate}

With the definition of the impedance network, we can associate the output voltage of the mycelium material at a specific position along the propagation direction of the electrical signal. By taking the limit with Kirchhoff's circuit laws \cite{irwin2020basic} on the infinitesimal segment, the voltage and current propagation along the mycelium can be represented as
\begin{equation}
\begin{aligned}
& \frac{\partial u(x, t)}{\partial x}=-\left(Ri(x, t)+ \frac{1}{C_1}  \int i(x, t) dt\right), \\
& \frac{\partial i(x, t)}{\partial x}=-\left(G u(x, t)+C_2 \frac{\partial u(x, t)}{\partial t}\right) ,
\end{aligned}
\label{eq:tel}
\end{equation}
Taking the derivative of the voltage equation concerning $x$ and substituting the time derivative of the current equation, we reach the following second-order partial differential equation for the voltage
\begin{equation}
\begin{aligned}
    \frac{\partial^2 u(x, t)}{\partial x^2}=&\left(RG+\frac{C_2}{C_1}\right)u(x, t)+RC_2\frac{\partial u(x, t)}{\partial t}\\
    &+\frac{G}{C_1}\int u(x, t) dt,
\end{aligned}
\label{eq:wave_v}
\end{equation}
Since Eq.~\ref{eq:wave_v} attains the form of a wave equation, its general solution is $ u\left(x,t\right)=U(x)e^{\iu\omega t}$, under the assumption of harmonic wave propagation. By substituting this ansatz of solution into Eq. \ref{eq:wave_v}, the frequency domain description of the voltage propagation is
\begin{equation}
    \frac{d^2U(x)}{dx^2}=\gamma^2 U(x), \quad \gamma=\sqrt{(R+\frac{1}{j \omega C_1})(G+j \omega C_2)},
    \label{eq:wave_vf}
\end{equation}
where $\gamma$ represents the complex propagation constant. The current equation $I(x)$ has the same form as the voltage equation in the frequency domain. Thus, the general solutions of $U(x)$ and $I(x)$ read
\begin{equation}
\begin{aligned}
    U(x)=U^{+}_{in} e^{-\gamma x}+U^{-}_{in} e^{\gamma x}, \\
    I(x)=\frac{U^{+}_{in}}{Z_0} e^{-\gamma x}-\frac{U^{-}_{in}}{Z_0} e^{\gamma x},
\end{aligned}
\end{equation}
with $U^{+}_{in}$ and $U^{-}_{in}$ denoting the forward and backward traveling voltages at the excitation point $x_0$, and
\begin{equation}
 Z_0=\sqrt{(R+\frac{1}{j \omega C_1})/(G+j\omega C_2)}
 \label{eq:characteristicImpendance}
\end{equation}
is the characteristic impedance of the impedance model. By applying the open circuit condition $I(L)=0$ with an infinite load resistance $R_l$, the solution of $U(x)$ can be simplified as
\begin{equation}
U(x)=U^{+}_{in} e^{-\gamma x}\left(1+e^{-2 \gamma(L-x)}\right).
\label{eq:ux}
\end{equation}
Therefore, the transfer function for the measured voltage at a given position $x$ along the sample can be expressed as
\begin{equation}
H(x, L)=\frac{U(x)}{U(0)}=\frac{e^{-\gamma x}\left(1+e^{-2 \gamma(L-x)}\right)}{\left(1+e^{-2 \gamma L}\right)}
\label{eq:frf}
\end{equation}
It is thus concluded that the propagation of the voltage source in the bounded mycelium material depends on the propagation constant at a given frequency and the location of the measurement point along the propagation direction. This indicates a dispersion relationship concerning the temporal and spatial directions simultaneously. Without properly characterizing this relationship, the voltage output of the fungal mycelium materials at a certain position may not yield meaningful results when adopting this for sensing purposes.

\subsection{Dispersion Analysis}
\label{sec:dispersion}

The dispersion relationship provides a general and fundamental description of the wave propagation characteristics of a linear and homogeneous material. Although the nature of the mycelium material at the micro level is intrinsically porous and inhomogeneous~\cite{islam2017morphology,olivero2023gradient}, this method can still provide valuable insight regarding the mycelium system on a macro level, if we assume the wavelength of the propagation signal is sufficiently larger than the size of the micro-level differences. In this case, the mycelium material can be regarded as the periodic arrangement of the infinite segment, as shown in the impedance network of the transmission line model provided in Fig.~\ref{fig:setup}.

As discussed in Sec. \ref{sec:impedance},  the propagation of the electrical signal $u\left(x,t\right)$ through the mycelium material along $0^\circ$ direction at frequency $\omega$ is defined as $ u\left(x,t\right)=U(x)e^{\iu\omega t}$. Regarding Eq. \ref{eq:ux}, we can let $L$ be infinite to cancel the boundary condition and obtain the plane wave representation of the traveling voltage signal as
\begin{equation}
    u\left(x,t\right)=U^{+}_{in} e^{-\gamma x}e^{\iu\omega t}=U^{+}_{in} e^{\iu\left(\omega t-kx\right)},
    \label{eq:planewave}
\end{equation}
where $k$ represents the wave number of the traveling electrical signal. The right-hand side of Eq.~\ref{eq:planewave} corresponds to a standard plane wave formulation within an infinite media. When considering the nondispersive wave propagation and material without loss, $k$ is a real number. However, it becomes a complex number when considering wave propagation in lossy or dispersive media. As documented in existing research~\cite{craster2012acoustic}, we utilize a complex wave number $k=k_r+\iu k_i$, where $k_r$ and $k_i$ are the real and imaginary wave numbers, respectively, to describe the dispersive and lossy characteristics of the mycelium material. From Eq.~\ref{eq:planewave}, it follows that $k(\omega)=-\iu \gamma$. This equation connects the impedance model discussed in Sec.~\ref{sec:impedance} to the dispersion relationship for the spatial and temporal characterization of the mycelium. Therefore, the analytical dispersion of the mycelium material reads
\begin{equation}
    k_r=\mathrm{Im}[\gamma], \quad k_i=-\mathrm{Re}[\gamma].
    \label{eq:k_vs_gamma}
\end{equation}
The real and imaginary parts of the propagation constant $\gamma$ represent the attenuation and phase constants of the propagation electromagnetic wave~\cite{pozar2021microwave}, which reversely correspond to the imaginary and real parts of the wave number~\cite{craster2012acoustic}. This indicates that the mycelium material's phase velocity and the attenuation of the propagation electrical signal are determined by the imaginary part and real part of the propagation constant $\gamma$, respectively.

\begin{figure}[!t]
\centering
\includegraphics[width=1\columnwidth,page=3]{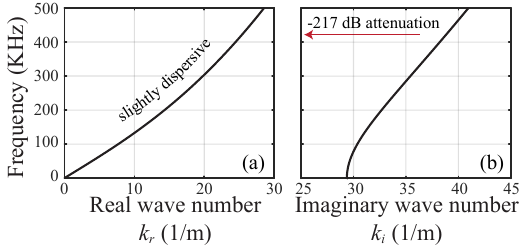}
\caption{Analytical wave numbers. (a) Real wave number $k_r$; (b) Imaginary wave number $k_i$.}
\label{fig:dis_theor}
\end{figure}

With the mean values of the fitted impedance parameters from Table~\ref{tab:para}, the real and imaginary wave numbers calculated by Eq.~\ref{eq:k_vs_gamma} are shown in Fig.~\ref{fig:dis_theor}. The form of the real wave number $k_r$ indicates that the propagation of the electrical signal is slightly dispersive; this, in turn, implies that the mycelium will generally keep the shape of the electrical signal for sensing purposes if the signal bandwidth is not too broad. The imaginary part wave number $k_i$ increases with frequency, exhibiting more substantial attenuation at higher frequencies and distances from the signal source. Further frequency increase leads to the dominance of $R$ and $C_2$. Real and imaginary wave numbers will follow $k_i=k_r=\sqrt{\omega RC_2/2}$, which hallmarks the reactive performance at high frequencies.

\begin{table}[h]
\centering
\caption{Fitted parameters per unit length}
\begin{tabular}{ccccc}
\hline
\begin{tabular}[c]{@{}c@{}} Parameter\end{tabular}
&  $R$ &  $C_1$ &  $C_2$ &  $G$ \\ \hline
\begin{tabular}[c]{@{}c@{}}Mean \\ value\end{tabular}
& 98 K$\Omega$      & 0.94 $\mu$F       & 7.6 nF       & 8.8e-3 $\Omega^{-1}$  \\ \hline
\begin{tabular}[c]{@{}c@{}}Standard \\ deviation\end{tabular}
& 6.5 K$\Omega$         & 0.16 $\mu$F        & 3.3 nF        & 2.4e-3 $\Omega^{-1}$
\\ \hline
\end{tabular}
\label{tab:para}
\end{table}

Besides the analytical method through the impedance model, we further utilize a purely statistical framework - the inhomogeneous wave correlation (IWC) method~\cite{van2018measuring,zhao2022graded}. The IWC provides estimates of the unknown complex wavenumbers by maximizing the correlation between a measured and an analytical wave field in the frequency domain.

If the continuous-time Fourier transform of the assumed traveling wave is
\begin{equation}
    V\left(x,k_r,k_i\right)=U^+_{in}e^{\left(-\iu k_rx+k_i x\right)}.
\end{equation}
and the Fourier transform of the measured response at frequency $\omega_0$ is notated as $U\left(x,\omega_0\right)$, the IWC function is defined as
\begin{equation}
\mathcal{C}\left(k_r, k_i\right)=
\frac{\left|\int U\left(x,\omega_0\right) \cdot V^*\left(x, k_r, k_i\right) \mathrm{d} x\right|}{\sqrt{\int\left|U\left(x, \omega_0\right)\right|^2 \mathrm{d} x \cdot \int\left|V\left(x, k_r, k_i\right)\right|^2 \mathrm{d} x}}
\end{equation}
where $*$ denotes complex conjugate. Due to the availability of digital recordings, the integrals are replaced with summations. The IWC function $\mathcal{C}$ typically has a well-defined maximum, at which the two transformations exhibit the highest correlation. This allows for an estimation of the complex wave number from
\begin{equation}
    \left(k_r,k_i\right)=\mathrm{agr} \, \underset{k_r,k_i}{\mathrm{max}} \mathcal{C}\left(k_r, k_i\right).
\end{equation}
It should be noted that the accurate estimation of the experimental wave numbers depends on the spatial resolution of the sampled points along a specific direction of the sample, at a fixed distance $d$ between two pairs of electrodes (Fig.~\ref{fig:setup}). Suppose the wavelength of the propagation signal is much smaller than $d$. In that case, the signal will be quickly damped due to the lossy nature of the mycelium, leading to inaccurate estimates of real wave numbers with insufficient and noisy data readings.

\begin{figure*}[!t]
\centering
\includegraphics[width=2\columnwidth,page=4]{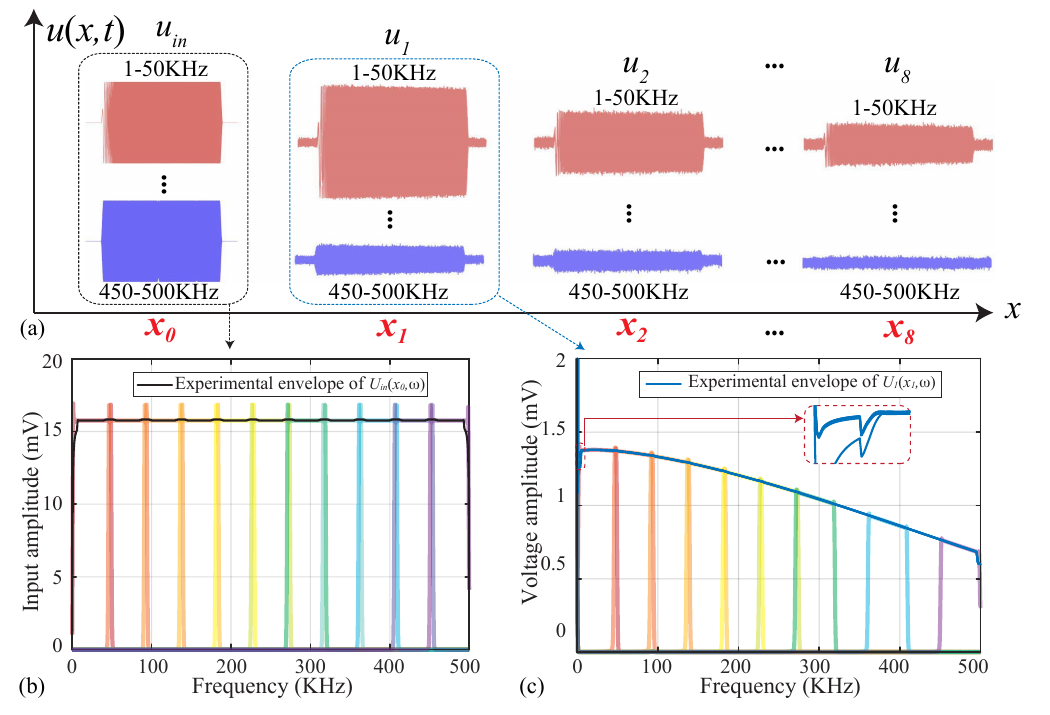}
\caption{Experimental signals and their frequency domain amplitudes. (a) The output electrical voltage along $0^\circ$ propagation. Each sampled position contains 11 segments of time domain sweeps up to 500 kHz; (b) The frequency domain amplitudes of the input electrical signal and the envelope of the 11 input segments; (c) The frequency domain amplitudes of the output electrical signal sampled at $x_1$ and their envelope curve. }
\label{fig:signal}
\end{figure*}

\begin{figure}[!t]
\centering
\includegraphics[width=1\columnwidth,page=5]{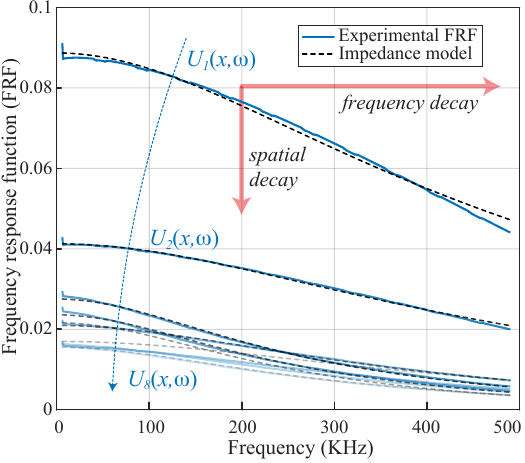}
\caption{Experimental and impedance model fitted frequency domain amplitudes sampled at $x_1$ to $x_8$.}
\label{fig:frf}
\end{figure}

\section{Experimental Results}
\label{sec:exp}
Based on the theoretical analysis above, the voltage excitation responses at different positions along different directions of the mycelium material have been recorded experimentally for frequency domain and dispersion analyses. These analyses not only provide how the electrical responses differ with the change of measurement position and frequency, but also provide insights into utilizing mycelium material as naturally embedded sensors in structures and buildings.

\subsection{Frequency Responses}

Limited by the buffer size of the Digilent devices, the experimental sweep signal is separated into 11 segments, spanning from 1 Hz to 500 kHz, with a constant time domain voltage amplitude of 1 V and a sampling frequency of 6.25 MHz. Each segment is executed five consecutive times, and the length of each data set is $65535$. Figure~\ref{fig:signal}(a) shows the voltage input, along with the measured responses at $x_0$, $x_1$, $x_2$, and $x_8$, for the first and the last segments. It can be seen that the responses are both frequency and location-dependent. Lower frequencies and closer proximity to the input source lead to larger voltage responses. Figure~\ref{fig:signal}(b) shows the input frequency spectrum of 11 segments.
Fig. \ref{fig:signal}(c) shows the frequency response of the measured voltage $u_1$ at $x_1$, which consists of 11 frequency response segments from separated input voltage excitations.
Due to the imperfect measurements, spline interpolations are used to fit the missing frequency domain data segments from 315 kHz to 365 kHz and from 405 kHz to 455 kHz in the frequency domain as shown in Fig. \ref{fig:signal}(c). The five envelope curves of the frequency responses measured at each position from $x_1$ to $x_8$ overlap with each other. An example of the overlapped envelope curves measured at $x_1$ is shown with the enlarged view in Fig. \ref{fig:signal}(c), demonstrating the experimental setup's robustness. In the following text, we only show the results calculated with the first excitation and disregard the frequency responses lower than 5 kHz and higher than 490 kHz to remove the influence of edge frequencies.

The experimental frequency response functions at positions $x_1$ to $x_8$ are demonstrated in Fig. \ref{fig:frf}.
There exist two decay trends with the increase of frequency and distance to the excitation source, which can be explained by the larger impedance amplitude at high frequencies and longer distance concerning the source position discussed in Sec. \ref{sec:impedance}. Therefore, by fitting the experimental frequency responses with the analytical transfer function proposed in Eq. \ref{eq:frf} at different positions, eight sets of parameters for the complex propagation constant $\gamma$ can be obtained. As the black dashed lines indicate, the fitted curves in Fig. \ref{fig:frf} show good agreement with the experimental frequency response functions.

\subsection{Dispersion Relationships}
\label{sec:exp_dis}
\begin{figure}[!t]
\centering
\includegraphics[width=1\columnwidth,page=6]{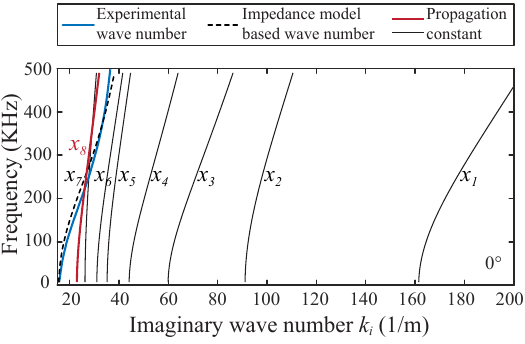}
\caption{ The imaginary wave numbers at $0^\circ$ degree direction along the mycelium. Experimental imaginary wave number $k_i$, impedance model-based wave number, and the real part of the propagation constant fitted by parameters measured at $x_1$ to $x_8$ are shown with different colors.}
\label{fig:para_fit}
\end{figure}

\begin{figure*}
\centering
\includegraphics[width=2\columnwidth,page=7]{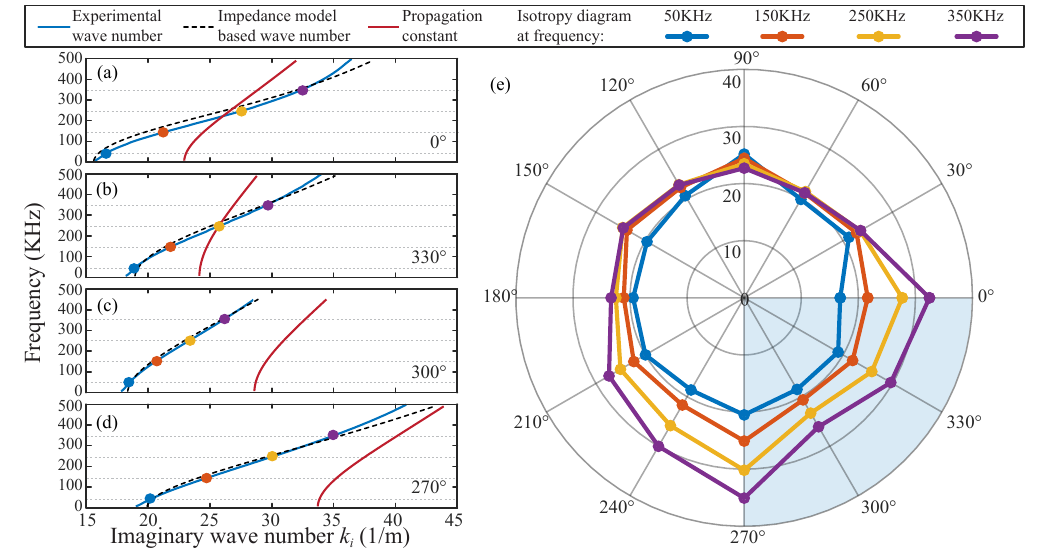}
\caption{Experimental imaginary wave numbers along different directions. (a) to (d): Experimental imaginary wave number $k_i$, impedance model based wave number, and the real part of the propagation constant along propagation direction $0^\circ$, $330^\circ$, $300^\circ$, and $270^\circ$, respectively. (e) The isotropy diagram of the mycelium material goes in different directions, and the four dotted lines correspond to frequency at 50 kHz, 150 kHz, 250 kHz, and 350 kHz, respectively.}
\label{fig:dis}
\end{figure*}

To reveal the spatial and temporal attenuation of the mycelium material, we have measured the frequency responses of the mycelium across different directions from $0^\circ$ to $330^\circ$. Each measurement follows the same procedures as the case of $0^\circ$ presented in the previous section. Due to disconnected sweep segments, only the amplitudes, rather than real and complex parts, of the frequency responses are used to compute the experimental and impedance model-based dispersion relationships. This leads to meaningful solutions for imaginary wave numbers. Since the amplitude in the frequency domain is real-valued, which does not include the phase information, the solutions of real wave numbers indicate that the phase velocities are disregarded. Fig. \ref{fig:para_fit} shows imaginary wave numbers calculated by the IWC method along the $0^\circ$ direction from experiments and the impedance-based model. The eight real part curves of the propagation constant $\gamma$ obtained by eight sets of the fitted parameters are also indicated with black and red lines.

Since the dispersion relationship cares for the electrical wave propagation with an infinite medium, the agreement between the propagation constant and the experimental imaginary wave number gradually increases with further points as shown with the propagation constant curves from $x_1$ to $x_8$ in Fig. \ref{fig:para_fit}.
Therefore, we utilize the parameters fitted from the farthest point at $x_8$ to approximate the propagation constant of an infinite mycelium media.
The mean values and standard deviations of the fitted  $R$, $C_1$, $C_2$, and $G$ per unit length spanning from $0^\circ$ to $330^\circ$ in total 12 directions are shown in Table \ref{tab:para}. It can be seen that these parameters are relatively stable across different directions with their standard deviations smaller than their mean values, which indicates the homogeneity of the mycelium material at the macro level.

Fig. \ref{fig:dis}(a) to (d) demonstrate the dispersion relationships of the mycelium material along $0^\circ$, $330^\circ$, $300^\circ$, and $270^\circ$ directions, which cover the blue shedding area in the isotropy diagram shown in
Fig. \ref{fig:dis}(e). For each direction, the experimental and impedance-based model dispersion curves are obtained by the IWC method with experimental and fitted amplitudes of each measured point along the propagation line. It can be seen that with the frequency increase, the imaginary wave number also increases, which reveals the attenuation of the propagating electrical signal along the propagation direction. This attenuation trend is not only frequency-dependent, as shown in Fig. \ref{fig:frf}, but also spatial-dependent due to the damped and dielectric nature of the living mycelium material, which indicates the limit of the size of the mycelium material when used as smart sensors. The real part of the propagation constant $\gamma$ is also presented as red lines in Fig. \ref{fig:dis}(a) to (d), whose growing trend generally agrees with the increase of the imaginary wave number. The discrepancy between the experimental wave number and the propagation constant suggested with Eq. \ref{eq:k_vs_gamma} could lie in: (1) The mismatch of the infinite media assumption of dispersion analysis and the finite size of the experimental mycelium sample; (2) Neglection of the frequency dependency of parameters of the impedance model; (3) The measurement noise under small values with high frequencies or distant sampled points.

The isotropy diagram is shown in Fig. \ref{fig:dis}(e) by computing the experimental imaginary wave numbers across different directions. Each point under a specific frequency is selected from the dispersion curves as shown by the dots with different colors in Fig. \ref{fig:dis}(a) to (d). It can be seen that the mycelium material is basically isotropic with a similar attenuation degree across different directions. The blue dotted line under 50 kHz indicates that the imaginary wave number is around 20, which means an exponential attenuation $e^{-20}\approx -174$dB per unit length. With the increase of frequency, the wave number also increases along all directions except $90^\circ$, which is regarded as an experimental outlier. For the isotropy diagrams at high frequencies, a deviation from isotropic distribution may exist due to indistinguishable and small measured values and noise.

\section{Discussion}
\label{sec:dis}

\begin{figure}[!t]
\centering
\includegraphics[width=1\columnwidth,page=8]{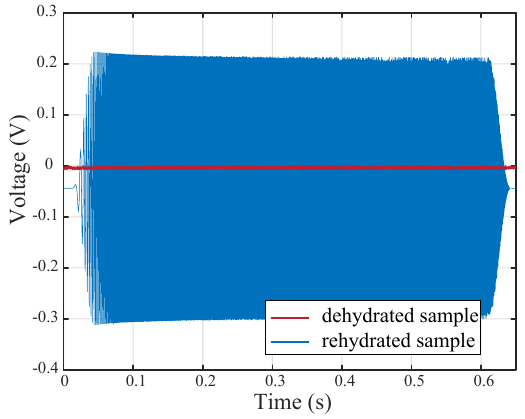}
\caption{Voltage measurements of the dehydrated and rehydrated sample.}
\label{fig:rehy}
\end{figure}

The impedance model and dispersion analysis offer analytical tools to describe the electrical response of mycelium, but also provide valuable insights for its use as a smart sensor or transducer. For propagation of high-frequency electrical signals (higher than 1 MHz), series resistance $R$ and parallel capacitance $C_2$ will dominate the dispersion relationship, leading to reactive responses. At lower frequencies, however, the series capacitance $C_1$ and conductance $G$ become increasingly significant, as shown in Fig. \ref{fig:dis_theor}. We further discuss the ionic conduction with free ions in the aqueous environment of the living mycelium, similar to conduction mechanisms in nerve tissues or across cell membranes  \cite{cragg1964conduction}. Unlike electromagnetic wave propagation, which can occur in dehydrated dielectric materials, ionic conduction in living mycelium depends on ion flow through channels and capacitive charging across membranes. As shown in Fig. \ref{fig:rehy}, we further dehydrated the mycelium sample using a microwave and measured the resulting voltage response of the dried mycelium sample under electrical excitation. The orange curve shows that the response is significantly reduced, indicating a low mobility of the free ions. This further implies that the electrical signal propagated in the living mycelium is carried by ionic mechanisms, since the electromagnetic wave can still propagate in dehydrated dielectric materials, as they do not rely on the presence of water or ions. We further rehydrated the mycelium, and the electrical signal can be recorded as a sweep signal, as shown in the blue curve. This ionic conduction mechanism reveals the dependency of the aqueous environment not only on the prosperity of the living mycelium, but also on the sensing abilities through electrical signal propagation.

    \begin{table*}[!t]
    \footnotesize
    \caption {Comparison with Existing Works} \label{compare}
    \centering
   \setlength{\tabcolsep}{3mm}{
   \renewcommand{\arraystretch}{3}
    \begin{tabular}{llllll}
    \toprule
    \textbf{Reference} & \textbf{Fungi investigated}& \textbf{Method used} & \textbf{Main observations} & \textbf{Potential application}     \\ \midrule
      Slayman and Slayman, 1962 \cite{slayman1962measurement}  &  \makecell[l]{Neurospora \\ crassa} & Microelectrodes      &   \makecell[l]{Existence of membrane \\ potential in fungi }     &   \makecell[l]{Environmental \\ sensing  }
    \\
      Horwitz et al., 1984 \cite{horwitz1984electric}  &   \makecell[l]{Trichoderma \\ harzianum }&
  \makecell[l]{Extracellular \\ microelectrodes}  &  \makecell[l]{Current and potential \\ responses under light stimuli}   &  \makecell[l]{Biophysical reaction \\ to light  }
    \\
      McGillivray and Gow, 1986 \cite{mcgillivray1986applied}  &   \makecell[l]{Neurospora \\ crassa}&
  \makecell[l]{Electrophoresis \\ machine}  &  \makecell[l]{Polarity of growth \\ under electrical field}   &  \makecell[l]{Controlled \\ fungi growth}
    \\
      Olsson and Hansson, 1995 \cite{olsson1995action}  &   \makecell[l]{Pleourotus \\ ostreatus}&
  \makecell[l]{Glass \\ microelectrodes}  &  \makecell[l]{Potential responses \\ under chemical stimuli}   &  \makecell[l]{Mycelium sensing \\ \& communication }
    \\
      \makecell[l]{Adamatzky et al., 2018 \cite{adamatzky2018spiking} \\ Adamatzky et al., 2021 \cite{adamatzkyFungalElectronics2021} \\ Adamatzky et al., 2022 \cite{adamatzky2022language} } &   \makecell[l]{Pleurotus djamor \\ Ganoderma resinaceum \\ Omphalotus nidiformis}&
  \makecell[l]{Subdermal needle \\ electrodes}  &  \makecell[l]{Voltage spiking \\ behaviors}   &  \makecell[l]{Voltage spikes for \\ communication \\ \& fungal language}
    \\
      Jones et al., 2024 \cite{jones2024fungal}  &   \makecell[l]{Curvularia \\ lunata}&
  \makecell[l]{Copper \\ electrode}  &  \makecell[l]{ Signal transmission \\ within fungi}   &  \makecell[l]{Mycelium as  \\ biosensors }
    \\
      Phillips et al., 2024 \cite{phillips2024electrical}  &   \makecell[l]{Pleurotus ostreatus\\ spawn}&
  \makecell[l]{Iridium needle \\ electrodes}  &  \makecell[l]{Signal transmission \\ under electrical stimuli}   &  \makecell[l]{ Bio-hybrid \\ computing systems}
    \\
      \textbf{This work}  &   \makecell[l]{Pleurotus ostreatus}&
  \makecell[l]{Iridium needle \\ electrodes}  &  \makecell[l]{Dispersion relationship \\ for electrical signal \\ propagation in mycelium}   &  \makecell[l]{Mycelium based \\ sensing systems }
    \\
    \bottomrule
    \end{tabular}}
    \end{table*}

    Research on electrical activities of fungal mycelium can be dated back to the early 1960s, during which Slayman and Slayman \cite{slayman1962measurement,slayman1976action} measured the membrane and action potentials of the fungus named Neurospora crassa. Their work serves as early biological proof for the existence of electrical activities in fungal mycelium. Since then, relevant research has been carried out in the past half century, as demonstrated in Table \ref{compare}. The early studies mainly focused on electrical responses, i.e., intracellular or extracellular action potentials, measured by microelectrodes. Taking the examples of the work by Horwitz et al. \cite{horwitz1984electric}, McGillivray and Gow \cite{mcgillivray1986applied}, and Olsson and Hansson \cite{olsson1995action}, they discovered the sensitivities of the fungal mycelium to light, electrical and chemical components, which facilitate the recent applications of mycelium-based sensors \cite{danninger2022myceliotronics} and robots \cite{mishra2024sensorimotor}. Parallel to the electrical measurements at the microscopic level, recent research also features the electrical measurements of mycelium at the macroscopic level. Adamatzky et al. \cite{adamatzky2018spiking,adamatzky2018towards,adamatzkyFungalElectronics2021,adamatzky2022language} demonstrated the "action-potential-like" spiking behaviors of the voltage reading of mycelium and their fruit bodies measured with subdermal needle electrodes. They claim that these spontaneous voltage spikes could refer to the fungal language. Although promising perspectives exist, there are criticisms about whether the measured spiking behaviors could be biologically derived \cite{blatt2024does}.

    Rather than claiming biological origin, recent studies by Jones et al. \cite{jones2024fungal} and Phillips et al. \cite{phillips2024electrical} discuss the electrical responses of mycelium materials as media for transmission of electrical signals. They revealed the dielectric characteristics of mycelium materials by measuring the transmitted signal under electrical voltage inputs across a broad bandwidth, indicating the applications of biohybrid computing and sensing systems. Compared with the studies mentioned above, the spatial and temporal characterization of the mycelium presented in this work does not focus on the biological origin of the action potentials or spiking behaviors of the mycelium. We base our study on the electrical responses of the mycelium system as a whole, which supports the conduction and propagation of electrical signals. Unlike local measurements of electrical responses \cite{jones2024fungal,phillips2024electrical}, this study emphasizes the propagation of electrical signals along a certain length of the mycelium material, which leads to the dependence of the propagation of the signal on distance and also the dispersion relationship of the mycelium material. With the proposed impedance network and dispersion analysis, this study offers insights and design guides for mycelium-based signal sensing and transmitting systems.

\section{Conclusion}
\label{sec:con}

In summary, this paper studies the spatial and temporal characterization of living mycelium through dispersion analysis. Rather than discussing the biological origin, we focus on the electrical responses under voltage stimulation of the living mycelium along the electrical signal propagation direction. An electrical impedance network model has been proposed for signal transmission along the mycelium material. With both dependencies on frequency and distance, the dispersion relationship of the living mycelium has been identified theoretically and experimentally. The electrical voltage responses indicate the attenuation of the signal propagation with increases of frequency and distance for the first time. This provides guidelines for the design of mycelium-based sensing and transmitting systems. Besides the dispersion relationship, the assumption of homogeneity and the dependency on aqueous environments have also been verified experimentally. The methods and results presented in this paper offer insights for the electrical signal propagation in living mycelium, which paves the way to integrate living mycelium as biosensors into mechanical systems for sustainable sensing and structural health monitoring applications.



\begin{thebibliography}{10}
\providecommand{\url}[1]{#1}
\csname url@samestyle\endcsname
\providecommand{\newblock}{\relax}
\providecommand{\bibinfo}[2]{#2}
\providecommand{\BIBentrySTDinterwordspacing}{\spaceskip=0pt\relax}
\providecommand{\BIBentryALTinterwordstretchfactor}{4}
\providecommand{\BIBentryALTinterwordspacing}{\spaceskip=\fontdimen2\font plus
\BIBentryALTinterwordstretchfactor\fontdimen3\font minus \fontdimen4\font\relax}
\providecommand{\BIBforeignlanguage}[2]{{%
\expandafter\ifx\csname l@#1\endcsname\relax
\typeout{** WARNING: IEEEtran.bst: No hyphenation pattern has been}%
\typeout{** loaded for the language `#1'. Using the pattern for}%
\typeout{** the default language instead.}%
\else
\language=\csname l@#1\endcsname
\fi
#2}}
\providecommand{\BIBdecl}{\relax}
\BIBdecl

\bibitem{coleineFungiAreKey2022}
C.~Coleine, J.~E. Stajich, and L.~Selbmann, ``Fungi are key players in extreme ecosystems,'' \emph{Trends in Ecology \& Evolution}, vol.~37, no.~6, pp. 517--528, Jun. 2022.

\bibitem{buffi2025electrical}
M.~Buffi, J.~M. Kelliher, A.~J. Robinson, D.~Gonzalez, G.~Cailleau, J.~A. Macalindong, E.~Frau, S.~Schintke, P.~S. Chain, C.~E. Stanley \emph{et~al.}, ``Electrical signaling in fungi: past and present challenges,'' \emph{FEMS Microbiology Reviews}, p. fuaf009, 2025.

\bibitem{mananSynthesisApplicationsFungal2021b}
S.~Manan, M.~W. Ullah, M.~{Ul-Islam}, O.~M. Atta, and G.~Yang, ``Synthesis and applications of fungal mycelium-based advanced functional materials,'' \emph{Journal of Bioresources and Bioproducts}, vol.~6, no.~1, pp. 1--10, Feb. 2021.

\bibitem{schrittSpentMushroomSubstrate2021}
H.~Schritt, S.~Vidi, and D.~Pleissner, ``Spent mushroom substrate and sawdust to produce mycelium-based thermal insulation composites,'' \emph{Journal of Cleaner Production}, vol. 313, p. 127910, Sep. 2021.

\bibitem{jonesLeatherlikeMaterialBiofabrication2021}
M.~Jones, A.~Gandia, S.~John, and A.~Bismarck, ``Leather-like material biofabrication using fungi,'' \emph{Nat Sustain}, vol.~4, no.~1, pp. 9--16, Jan. 2021.

\bibitem{jones2017mycelium}
M.~Jones, T.~Huynh, C.~Dekiwadia, F.~Daver, and S.~John, ``Mycelium composites: a review of engineering characteristics and growth kinetics,'' \emph{Journal of Bionanoscience}, vol.~11, no.~4, pp. 241--257, 2017.

\bibitem{jones2020engineered}
M.~Jones, A.~Mautner, S.~Luenco, A.~Bismarck, and S.~John, ``Engineered mycelium composite construction materials from fungal biorefineries: A critical review,'' \emph{Materials \& Design}, vol. 187, p. 108397, 2020.

\bibitem{adamatzky2023fungal}
A.~Adamatzky, \emph{Fungal machines: Sensing and computing with fungi}.\hskip 1em plus 0.5em minus 0.4em\relax Springer Nature, 2023, vol.~47.

\bibitem{mishra2024sensorimotor}
A.~K. Mishra, J.~Kim, H.~Baghdadi, B.~R. Johnson, K.~T. Hodge, and R.~F. Shepherd, ``Sensorimotor control of robots mediated by electrophysiological measurements of fungal mycelia,'' \emph{Science Robotics}, vol.~9, no.~93, p. eadk8019, 2024.

\bibitem{schyck2024harnessing}
S.~Schyck, P.~Marchese, M.~Amani, M.~Ablonczy, L.~Spoelstra, M.~Jones, Y.~Bathaei, A.~Bismarck, and K.~Masania, ``Harnessing fungi signaling in living composites,'' \emph{Global Challenges}, vol.~8, no.~8, p. 2400104, 2024.

\bibitem{gow1984transhyphal}
N.~A. Gow, ``Transhyphal electrical currents in fungi,'' \emph{Microbiology}, vol. 130, no.~12, pp. 3313--3318, 1984.

\bibitem{gow1995electric}
N.~Gow and B.~Morris, ``The electric fungus,'' \emph{Botanical Journal of Scotland}, vol.~47, no.~2, pp. 263--277, 1995.

\bibitem{slayman1962measurement}
C.~L. Slayman and C.~W. Slayman, ``Measurement of membrane potentials in neurospora,'' \emph{Science}, vol. 136, no. 3519, pp. 876--877, 1962.

\bibitem{slayman1976action}
C.~L. Slayman, W.~S. Long, and D.~Gradmann, ``“action potentials” in neurospora crassa, a mycelial fungus,'' \emph{Biochimica et Biophysica Acta (BBA)-Biomembranes}, vol. 426, no.~4, pp. 732--744, 1976.

\bibitem{adamatzky2018spiking}
A.~Adamatzky, ``On spiking behaviour of oyster fungi pleurotus djamor,'' \emph{Scientific reports}, vol.~8, no.~1, p. 7873, 2018.

\bibitem{adamatzkyFungalElectronics2021}
A.~Adamatzky, P.~Ayres, A.~E. Beasley, A.~Chiolerio, M.~M. Dehshibi, A.~Gandia, E.~Albergati, R.~Mayne, A.~Nikolaidou, N.~Roberts \emph{et~al.}, ``Fungal electronics,'' \emph{Biosystems}, vol. 212, p. 104588, 2022.

\bibitem{adamatzky2022language}
A.~Adamatzky, ``Language of fungi derived from their electrical spiking activity,'' \emph{Royal Society Open Science}, vol.~9, no.~4, p. 211926, 2022.

\bibitem{dehshibiElectricalActivityFungi2021}
M.~M. Dehshibi and A.~Adamatzky, ``Electrical activity of fungi: {{Spikes}} detection and complexity analysis,'' \emph{Biosystems}, vol. 203, p. 104373, May 2021.

\bibitem{danninger2022myceliotronics}
D.~Danninger, R.~Pruckner, L.~Holzinger, R.~Koeppe, and M.~Kaltenbrunner, ``Myceliotronics: Fungal mycelium skin for sustainable electronics,'' \emph{Science Advances}, vol.~8, no.~45, p. eadd7118, 2022.

\bibitem{ganzeboom2024fungal}
S.~Ganzeboom, B.~Zhao, V.~Dertimanis, and E.~Chatzi, ``Fungal circuitry: mycelium as a living sensor for smart structures,'' in \emph{Sensors and Smart Structures Technologies for Civil, Mechanical, and Aerospace Systems 2024}, vol. 12949.\hskip 1em plus 0.5em minus 0.4em\relax SPIE, 2024, pp. 106--115.

\bibitem{phillips2023electrical}
N.~Phillips, A.~Gandia, and A.~Adamatzky, ``Electrical response of fungi to changing moisture content,'' \emph{Fungal Biology and Biotechnology}, vol.~10, no.~1, p.~8, 2023.

\bibitem{olsson1995action}
S.~Olsson and B.~Hansson, ``Action potential-like activity found in fungal mycelia is sensitive to stimulation,'' \emph{Naturwissenschaften}, vol.~82, pp. 30--31, 1995.

\bibitem{nikolaidou2023responsive}
A.~Nikolaidou, N.~Phillips, M.-A. Tsompanas, and A.~Adamatzky, ``Responsive fungal insoles for pressure detection,'' \emph{Scientific Reports}, vol.~13, no.~1, p. 4595, 2023.

\bibitem{mcgillivray1986applied}
A.~M. McGILLIVRAY and N.~A. Gow, ``Applied electrical fields polarize the growth of mycelial fungi,'' \emph{Microbiology}, vol. 132, no.~9, pp. 2515--2525, 1986.

\bibitem{phillips2024electrical}
N.~Phillips, R.~Weerasekera, N.~Roberts, A.~Gandia, and A.~Adamatzky, ``Electrical signal transfer characteristics of mycelium-bound composites and fungal fruiting bodies,'' \emph{Fungal Ecology}, vol.~71, p. 101358, 2024.

\bibitem{horwitz1984electric}
B.~A. Horwitz, M.~H. Weisenseel, A.~Dorn, and J.~Gressel, ``Electric currents around growing trichoderma hyphae, before and after photoinduction of conidiation,'' \emph{Plant physiology}, vol.~74, no.~4, pp. 912--916, 1984.

\bibitem{fukasawa2024electrical}
Y.~Fukasawa, D.~Akai, T.~Takehi, and Y.~Osada, ``Electrical integrity and week-long oscillation in fungal mycelia,'' \emph{Scientific Reports}, vol.~14, no.~1, p. 15601, 2024.

\bibitem{jones2024fungal}
R.~M. Jones, R.~W. Reynolds, A.~K. Thurston, and R.~A. Barbato, ``Fungal tissue as a medium for electrical signal transmission: A baseline assessment with melanized fungus curvularia lunata,'' \emph{IEEE Journal of Electromagnetics, RF and Microwaves in Medicine and Biology}, 2024.

\bibitem{pozar2021microwave}
D.~M. Pozar, \emph{Microwave engineering: theory and techniques}.\hskip 1em plus 0.5em minus 0.4em\relax John wiley \& sons, 2021.

\bibitem{stupin2021bioimpedance}
D.~D. Stupin, E.~A. Kuzina, A.~A. Abelit, A.~K. Emelyanov, D.~M. Nikolaev, M.~N. Ryazantsev, S.~V. Koniakhin, and M.~V. Dubina, ``Bioimpedance spectroscopy: basics and applications,'' \emph{ACS Biomaterials Science \& Engineering}, vol.~7, no.~6, pp. 1962--1986, 2021.

\bibitem{schwan1992linear}
H.~Schwan, ``Linear and nonlinear electrode polarization and biological materials,'' \emph{Annals of biomedical engineering}, vol.~20, pp. 269--288, 1992.

\bibitem{dean2008electrical}
D.~Dean, T.~Ramanathan, D.~Machado, and R.~Sundararajan, ``Electrical impedance spectroscopy study of biological tissues,'' \emph{Journal of electrostatics}, vol.~66, no. 3-4, pp. 165--177, 2008.

\bibitem{seyedi2013survey}
M.~Seyedi, B.~Kibret, D.~T. Lai, and M.~Faulkner, ``A survey on intrabody communications for body area network applications,'' \emph{IEEE Transactions on Biomedical Engineering}, vol.~60, no.~8, pp. 2067--2079, 2013.

\bibitem{callejon2011study}
M.~A. Callejon, L.~M. Roa, J.~Reina-Tosina, and D.~Naranjo-Hernandez, ``Study of attenuation and dispersion through the skin in intrabody communications systems,'' \emph{IEEE Transactions on Information Technology in Biomedicine}, vol.~16, no.~1, pp. 159--165, 2011.

\bibitem{wu2021electrical}
Y.~Wu, F.~F. Hanzaee, D.~Jiang, R.~H. Bayford, and A.~Demosthenous, ``Electrical impedance tomography for biomedical applications: Circuits and systems review,'' \emph{IEEE Open Journal of Circuits and Systems}, vol.~2, pp. 380--397, 2021.

\bibitem{craster2012acoustic}
R.~V. Craster and S.~Guenneau, \emph{Acoustic metamaterials: negative refraction, imaging, lensing and cloaking}.\hskip 1em plus 0.5em minus 0.4em\relax Springer Science \& Business Media, 2012, vol. 166.

\bibitem{de2024localized}
J.~M. De~Ponti, X.~Zhao, L.~Iorio, T.~Maggioli, M.~Colangelo, B.~Davaji, R.~Ardito, R.~V. Craster, and C.~Cassella, ``Localized topological states beyond fano resonances via counter-propagating wave mode conversion in piezoelectric microelectromechanical devices,'' \emph{Nature Communications}, vol.~15, no.~1, p. 9617, 2024.

\bibitem{zhao2024nonlinear}
B.~Zhao, H.~R. Thomsen, X.~Pu, S.~Fang, Z.~Lai, B.~Van~Damme, A.~Bergamini, E.~Chatzi, and A.~Colombi, ``A nonlinear damped metamaterial: Wideband attenuation with nonlinear bandgap and modal dissipation,'' \emph{Mechanical Systems and Signal Processing}, vol. 208, p. 111079, 2024.

\bibitem{zhao2023circuit}
B.~Zhao, J.~Qiu, and J.~Liang, ``Circuit solutions toward broadband piezoelectric energy harvesting: An impedance analysis,'' \emph{IEEE Transactions on Circuits and Systems I: Regular Papers}, 2023.

\bibitem{van2018measuring}
B.~Van~Damme and A.~Zemp, ``Measuring dispersion curves for bending waves in beams: a comparison of spatial fourier transform and inhomogeneous wave correlation,'' \emph{Acta Acustica united with Acustica}, vol. 104, no.~2, pp. 228--234, 2018.

\bibitem{zhao2022graded}
B.~Zhao, H.~R. Thomsen, J.~M. De~Ponti, E.~Riva, B.~Van~Damme, A.~Bergamini, E.~Chatzi, and A.~Colombi, ``A graded metamaterial for broadband and high-capability piezoelectric energy harvesting,'' \emph{Energy Conversion and Management}, vol. 269, p. 116056, 2022.

\bibitem{grownInnovativeMycelium}
``{I}nnovative {M}ycelium {P}ackaging for {A}ll {K}inds of {P}roducts --- grown.bio,'' \url{https://www.grown.bio/}, 2025.

\bibitem{foster2019dielectric}
K.~R. Foster and H.~P. Schwan, ``Dielectric properties of tissues,'' \emph{CRC handbook of biological effects of electromagnetic fields}, pp. 27--96, 2019.

\bibitem{mcadams1995tissue}
E.~McAdams and J.~Jossinet, ``Tissue impedance: a historical overview,'' \emph{Physiological measurement}, vol.~16, no.~3A, p.~A1, 1995.

\bibitem{irwin2020basic}
J.~D. Irwin and R.~M. Nelms, \emph{Basic engineering circuit analysis}.\hskip 1em plus 0.5em minus 0.4em\relax John Wiley \& Sons, 2020.

\bibitem{islam2017morphology}
M.~R. Islam, G.~Tudryn, R.~Bucinell, L.~Schadler, and R.~Picu, ``Morphology and mechanics of fungal mycelium,'' \emph{Scientific reports}, vol.~7, no.~1, p. 13070, 2017.

\bibitem{olivero2023gradient}
E.~Olivero, E.~Gawronska, P.~Manimuda, D.~Jivani, F.~Z. Chaggan, Z.~Corey, T.~S. de~Almeida, J.~Kaplan-Bie, G.~McIntyre, O.~Wodo \emph{et~al.}, ``Gradient porous structures of mycelium: A quantitative structure--mechanical property analysis,'' \emph{Scientific Reports}, vol.~13, no.~1, p. 19285, 2023.

\bibitem{cragg1964conduction}
B.~Cragg and P.~Thomas, ``The conduction velocity of regenerated peripheral nerve fibres,'' \emph{The Journal of physiology}, vol. 171, no.~1, p. 164, 1964.

\bibitem{adamatzky2018towards}
A.~Adamatzky, ``Towards fungal computer,'' \emph{Interface focus}, vol.~8, no.~6, p. 20180029, 2018.

\bibitem{blatt2024does}
M.~R. Blatt, G.~K. Pullum, A.~Draguhn, B.~Bowman, D.~G. Robinson, and L.~Taiz, ``Does electrical activity in fungi function as a language?'' \emph{Fungal Ecology}, vol.~68, p. 101326, 2024.

\end{thebibliography}
\end{document}